 \documentclass[final,3p,times,twocolumn]{elsarticle}

\usepackage{amssymb}
\usepackage{amsmath}
\usepackage{color}
\usepackage{graphicx}
\usepackage{subcaption}

\journal{Physics Letter B}

\begin{document}

\begin{frontmatter}


 \author{Jingwei Liu\corref{cor1}\fnref{label1}}
 \ead{jingweil@arizona.edu}
 \cortext[cor1]{Corresponding author}

\title{A Truncated Primordial Power Spectrum and \\
Its Impact on B-Mode Polarization}

 \author[label2]{Fulvio Melia}
 \affiliation[label1]{organization={Department of Physics, The University of Arizona},
             state={AZ},
             postcode={85721},
             country={USA}}

 \affiliation[label2]{organization={Department of Physics, The Applied Math Program,\\ and Department of
Astronomy, The University of Arizona},
             state={AZ},
             postcode={85721},
             country={USA}}

\begin{abstract}
The absence of large-angle correlations in the temperature of the cosmic 
microwave background (CMB), confirmed by three independent satellite missions, 
creates significant tension with the standard model of cosmology. Previous 
work has shown, however, that a truncation, $k_{\rm min}$, of the primordial 
power spectrum comprehensively resolves the anomaly and the missing power 
at $\ell\lesssim 5$ (the low multipoles). Since this cutoff is consistent 
with the hypothesized delay of inflation well beyond the Planck time, we
are strongly motivated to consider its possible impact on other observational 
signatures. In this {\it Letter}, we analyze and predict its influence on 
the most revealing probe awaiting measurement by upcoming missions---the 
B-mode polarization of the CMB, whose accurate determination should greatly
impact the inflationary picture. We highlight the quantitative power of this
discriminant by specifically considering the LiteBIRD mission, predicting
the effect of $k_{\rm min}$ on both the angular power spectrum and the 
angular correlation function of the B-mode, for a range of tensor-to-scalar 
ratios, $r$. While its impact on the latter appears to be negligible, 
$k_{\rm min}$ should have a very pronounced effect on the former. We show 
that for $r=0.036$, $k_{\rm min}$'s impact on $C_{\ell}^{BB}$ at low 
$\ell$'s should be easily detectable by LiteBIRD, but will be 
largely hidden by the total uncertainty of the measurement if $r\lesssim 0.02$.
\end{abstract}




\end{frontmatter}


\section{Introduction}
Inflationary cosmology \cite{Starobinskii:1979,Kazanas:1980,Guth:1981,Linde:1982}, 
offering insights into the early Universe's rapid expansion, may overcome several 
difficulties with the standard model, including the cosmic microwave background's 
(CMB) temperature horizon problem and the scarcity of magnetic monopoles. Many 
details regarding inflation, however, including the specifics of the driving field 
and the onset of the rapid expansion, remain elusive.

These uncertainties have been compounded since the {\it Planck}-2013 data release 
\cite{Planck:2014}, which shifted the focus from simple, chaotic inflation to models 
featuring plateau-like potentials, removing the possibility of matching
the Universe's energy density to the Planck scale at the Planck time, $t_{\rm Pl}$ 
\cite{Ijjas:2013,Ijjas:2014}. As we now understand, however, a possibly related 
anomaly emerges from the observed lack of large-angle correlations in the CMB,
 confirmed by three independent satellite missions: the Cosmic 
Background Explorer (COBE) \cite{Hinshaw:1996}, the Wilkinson Microwave Anisotropy 
Probe (WMAP) \cite{Bennett:2003} and the Planck mission \cite{Planck:2014,PlanckVI:2020}.
This issue is not effectively resolved by cosmic variance, which assigns an 
unacceptably low probability to the observed Universe \cite{Copi:2009}.

 Instead, a comprehensive analysis of the {\it Planck} data suggests 
that the absence of large-angle correlations in the CMB can be explained in terms 
of a non-zero minimum wavenumber, 
$k_{\rm min}$, in the primordial power spectrum, $P(k)$ \cite{MeliaLopez:2018}. 
Previous work has shown that such a truncation would naturally arise if the 
de Sitter expansion began at a time $t_{\rm init}\gg t_{\rm Pl}$ \cite{LiuMelia:2020}.

We thus appear to be witnessing an interesting confluence of modern inflationary
theories and recent observations both pointing to a delayed onset of de Sitter
expansion, well past $t_{\rm Pl}$, producing a non-zero minimum wavenumber, 
$k_{\rm min}$, in $P(k)$.

In this {\it Letter}, we focus on the broader impact of this cutoff on the 
observational signatures created by inflation. Starting with the role of 
$k_{\rm min}$ in mitigating the TT discrepancies, we extend the analysis to
gauge its impact on upcoming observations of the CMB polarization. If the
latter confirm its reality, the existence of such a cutoff would represent 
one of the most important, if yet unexplored, features of the primordial 
power spectrum.  For example, a nonzero value of $k_{\rm min}$ affects the 
angular correlation function of the B-mode polarization, whose low-$\ell$ 
portion is predominantly generated by tensor fluctuations.

We utilize the CAMB code \cite{Lewis:2000} with the latest {\it Planck} 
results \cite{PlanckVI:2020} to compute the angular power spectrum and 
corresponding angular correlation function of the polarization signal 
for a range of $k_{\rm min}$ values. To provide a quantitative measure 
of how this cutoff should impact future observations, we evaluate the 
expected observational signal relative to the sensitivity of LiteBIRD 
\cite{LiteBIRD:2022}. In doing so, we highlight the significance of 
the tensor-to-scalar ratio $r$ in modulating the impact of $k_{min}$ 
within the conventional inflationary paradigm.

\section{A Truncated Primordial Power Spectrum}\label{truncated}
\subsection{Background}
The primordial power spectrum was generated when the quantum fluctuations
in the inflaton field crossed the Hubble horizon. During inflation, the 
Hubble radius, $R_{\rm h} \equiv c/H(t)$, remained nearly constant, while 
the mode wavelengths, $\lambda_k = 2\pi a(t)/k$, grew at a rate proportional
to the expansion factor $a(t)$. Modes with different $k$'s crossed $R_{\rm h}$ 
at different times \cite{Bardeen:1983}, so a de Sitter expansion beginning at 
a particular time, $t_{\rm init}$, would have produced a largest mode 
corresponding to $k_{\rm min}$ \cite{LiuMelia:2020}. 

The angular power spectra for the temperature and polarization in the CMB 
largely depend on $P(k)$ \cite{Zaldarriaga:1997,Kosowsky:1996,Polnarev:1985}:
\begin{equation}
C_{\{T,E\}\ell}^{(S)} = 4\pi^2 \int_{k_{\rm min}}^{\infty} \, dk \, k^2 
P_{\phi}(k) \left[\Delta_{\{T,E\}\ell}^{(S)}(k)\right]^2,
\end{equation}
\begin{equation}
C_{\{T,E,B\}\ell}^{(T)} = 4\pi^2 \int_{k_{\rm min}}^{\infty} \, dk \, 
k^2 P_{h}(k) \left[\Delta_{\{T,E,B\}\ell}^{(T)}(k)\right]^2.
\end{equation}
In these equations, \(T\), \(E\), and \(B\) refer to the temperature 
fluctuations, the E-mode polarization, and B-mode polarization, respectively. 
The quantities $(S)$ and $(T)$ denote scalar and tensor modes, and
\(P_\phi\) and \(P_h\) represent the primordial scalar and tensor power 
spectra, while \(\Delta\) is the transfer function. 

The angular correlation function, $C(\theta)$, may be written in
terms of the angular power spectrum 
\cite{Zaldarriaga:1997,Kosowsky:1996,Polnarev:1985,Yoho:2015}:
\begin{equation}
C_{TT}(\theta) = \frac{1}{4\pi} \sum_{\ell=2}^{\infty} (2\ell + 1) 
C_{\ell}^{TT} P_\ell(\cos\theta),
\end{equation}
\begin{equation}\label{CthetaEE}
C_{\hat{E}\hat{E}}(\theta) = \frac{1}{4\pi} \sum_{\ell=2}^{\infty} 
(2\ell + 1) \frac{(\ell+2)!}{(\ell-2)!} C_{\ell}^{EE} P_\ell(\cos\theta),
\end{equation}
\begin{equation}\label{CthetaBB}
C_{\hat{B}\hat{B}}(\theta) = \frac{1}{4\pi} \sum_{\ell=2}^{\infty} 
(2\ell + 1) \frac{(\ell+2)!}{(\ell-2)!} C_{\ell}^{BB} P_\ell(\cos\theta),
\end{equation}
where $P_\ell(\cos\theta)$ are the Legendre polynomials. We shall here 
introduce the quantities $\hat{B}$ and $\hat{E}$ to represent the angular 
correlation function of the polarization, which we prefer over 
the more commonly used $E$ and $B$ modes. The latter are non-local 
quantities requiring an all-sky integral for extraction. In contrast, 
$\hat{B}$ and $\hat{E}$ are designed to be computable from the $Q$ 
and $U$ modes (both observable quantities) by applying local 
spin-raising and spin-lowering operators \cite{Zaldarriaga:1997,Kosowsky:1996}:
\begin{equation}
\hat{E}(\hat{n}) = -\frac{i}{2} \Big[ \bar{\eth}^2 (Q(\hat{n}) + 
iU(\hat{n})) - \eth^2 (Q(\hat{n}) - iU(\hat{n})) \Big]\quad
\end{equation}
\begin{equation}
\hat{B}(\hat{n}) = \frac{1}{2} \Big[ \bar{\eth}^2 (Q(\hat{n}) + 
iU(\hat{n})) + \eth^2 (Q(\hat{n}) - iU(\hat{n})) \Big]\;.\quad
\end{equation}

A non-zero \(k_{\rm min}\) affects all the \(C_l\) values and 
\(C(\theta)\) at every angle. We shall reproduce this result 
below and demonstrate the dramatic improvement of the model
fit compared to the analogous situation with $k_{\rm min}=0$. 
Given that \(k_{\rm min}\) is primarily a large-scale feature, 
however, its impact is mainly felt at large angles and low 
$\ell$'s, where it suppresses the values of \(C_\ell\). 

The outcome of calculating \(C(\theta)\) is quite different for 
the temperature and polarization cases. As demonstrated below,
the suppression of \(C_\ell\) for the temperature at low $\ell$'s
induces a notable shift in the \(C_{TT}(\theta)\) curve. 
But the situation changes significantly for the polarization. 
Due to the additional \(\frac{(\ell+2)!}{(\ell-2)!}\) factor, 
\(C_{EE,BB}(\theta)\) is dominated by the large-$\ell$ 
terms \cite{Yoho:2015}, thereby weakening the impact of 
\(k_{\rm min}\).

The effect of $k_{\rm min}$ on \(C_{TT}(\theta)\) was one of the 
principal reasons for its introduction. Previous work 
\cite{MeliaLopez:2018} demonstrated that the {\it Planck} data 
decidedly rule out a zero $k_{\rm min}$ at a confidence level 
exceeding $8\sigma$. The value of the cutoff is instead measured
to be
\begin{equation}
k_{\rm min} = (3.14 \pm 0.36) \times 10^{-4} \quad {\rm Mpc}^{-1},\label{eq:kmin}
\end{equation}
for a last scattering surface at $z=1080$. We shall adopt this 
cutoff throughout this work. 

Its viability for resolving the large-angle anomalies in the CMB 
was reinforced by a second study \cite{Melia:2021b} focused on the 
TT angular power spectrum. This subsequent work demonstrated a clear
improvement of the model fit at $\ell \leq 5$, with an optimized cutoff
in this case at $k_{\rm min} = (2.04^{+1.4}_{-0.79}) \times 10^{-4} \;
{\rm Mpc}^{-1}$, consistent with the previous value to within $1\sigma$.

To be clear, these two earlier papers \cite{MeliaLopez:2018,Melia:2021b}
analyzed two separate features of the CMB anisotropies. The first was
solely focused on the angular correlation function, while the second 
addressed the low power in the low-$\ell$ multipole components. 
The strong evidence ($\gtrsim 8\sigma$) against a zero value for $k_{\rm min}$ 
was based purely on the impact of such a cutoff on what is essentially
the standard model. The inferred $k_{\rm min}$ optimized the fit to the 
angular correlation function. Of course, this outcome does not necessarily
imply that the standard model is ruled out at this confidence level. It merely
states that $k_{\rm min}=0$ is ruled out, whether the underlying model is 
$\Lambda$CDM or something else. In this regard, we point out that the
`new' chaotic inflationary paradigm itself requires a delayed initiation 
to inflation, which necessarily implies that $k_{\rm min}\not=0$, unlike
the original (or classical) version of the model in which the 
fluctuations were believed to have originated below the Planck scale, 
thereby implying that $k_{\rm min}$ is effectively zero \citep{LiuMelia:2024}.

Given the impact of such a strong result, we here summarize the procedure
followed in \cite{MeliaLopez:2018} to arrive at this $8 \sigma$ confidence
level. The points in the observed angular correlation function $C(\theta)$ 
are highly correlated. This needs to be taken into account in the statistical 
analysis, as described in that earlier publication. The optimized value
of $k_{\rm min}$ is obtained by examining the distribution of mock CMBR
catalogs (using standard cosmology with the angular correlation function
$C[\theta]$). Its uncertainty is derived from the r.m.s. value within which
one finds $68\,\%$ of the possible outcomes. This error is actually larger 
than the value one would infer from a simple $\chi^2$ fitting, which would
ignore the correlations. In other words, the estimate of $8 \sigma$
is atually the most conservative limit one may get when analyzing
the angular correlation function on its own.

Our second publication \citep{Melia:2021b} complemented this earlier work
by considering the low power anomaly associated with the low-$\ell$
multipoles. Its purpose was to see whether or not the $k_{\rm min}$
inferred from $C(\theta)$ also solves this independent problem with
the traditional $CMB$ analysis. The simple answer is `yes,' but
the confidence level associated with a non-zero $k_{\rm min}$ based
solely on the low power argument is weaker: $\sim 2.6 \sigma$. Nevertheless, 
these two analyses together point to a compelling scenario, in which a 
non-zero $k_{\rm min}$ can easily solve both anomalies with the value 
quoted in Equation~\ref{eq:kmin}.

Crucially, this second study also confirmed that the inclusion of 
$k_{\rm min}$ as one of the free parameters in the standard model 
creates no noticeable change to the other variables, whose previous 
optimization produced the well-known, impressive fit to the angular 
power spectrum at $\ell\gg 10$. For simplicity, we shall therefore 
simply adopt the {\it Planck}-concordance values of all the cosmological 
parameters except for $k_{\rm min}$.

\subsection{Methodology}
In spite of these compelling results with the TT anisotropies, however, 
there are two principal reasons for focusing on the CMB polarization 
going forward. First, forthcoming observations are likely to provide 
measurements of the polarization precise enough to distinguish between 
scenarios with and without $k_{\rm min}$. Second, B-mode polarization 
is a tell-tale signature of tensor modes, which are masked by scalar-mode 
signals in the TT and EE anisotropies. This information is critical for 
improving our understanding of inflation and the early Universe. Findings 
related to the TT anisotropies merely confirm the existence of 
$k_{\rm min}$ for the primordial scalar-mode power spectrum. But if 
$k_{\rm min}$ emerges due to the delayed commencement of inflation, it 
should also be present in the primordial tensor spectrum. This can only 
be identified through the analysis of B-mode polarization. 

\begin{figure}[ht]
\centering
\includegraphics[width=\columnwidth]{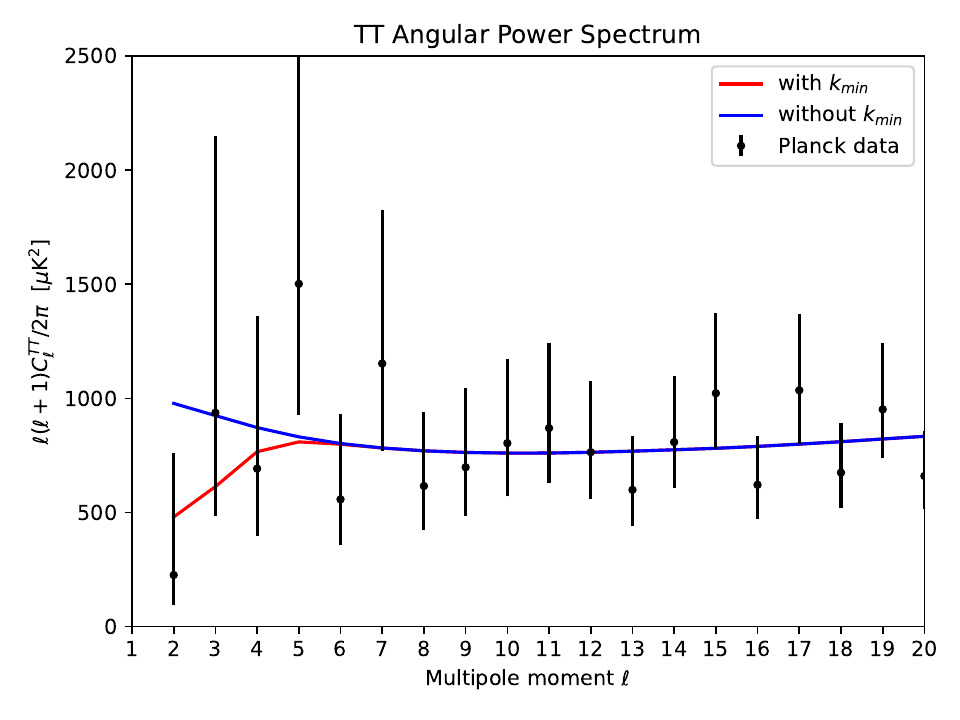}
\caption{Calculated TT angular power spectrum with (red)
and without (blue) the cutoff $k_{\rm min}$, in comparison
with the {\it Planck} data.}
\label{fig:ttcl}
\end{figure}

All of our calculations are carried out using the \texttt{CAMB} 
code (version 1.4.0) with the most recent {\it Planck} results 
as its default parameters \cite{Lewis:2000}. The 
cutoff is introduced to the default primordial power spectrum as follows 
\begin{equation}
P^{(S)}(k) = 
\begin{cases} 
0 & \text{for } k < k_{\rm min}\,, \\
A_{\rm s} \cdot 
\left(\frac{k}{k_{\rm pivot}}\right)^{n_{\rm s} - 1} & \text{for } k \geq k_{\rm min}\,.
\end{cases}
\label{eq:Pk}
\end{equation}
in terms of the amplitude, $A_{\rm s}$, of the primordial power spectrum, the spectral index 
$n_{\rm s}$ of the scalar fluctuations, and the pivot scale $k_{\rm pivot}=0.05\;{\rm Mpc}^{-1}$. 
The rest of the parameters used in these computations are derived from the most recent 
{\it Planck} optimizations, as discussed above. 

\begin{figure}[ht]
\centering
\includegraphics[width=\columnwidth]{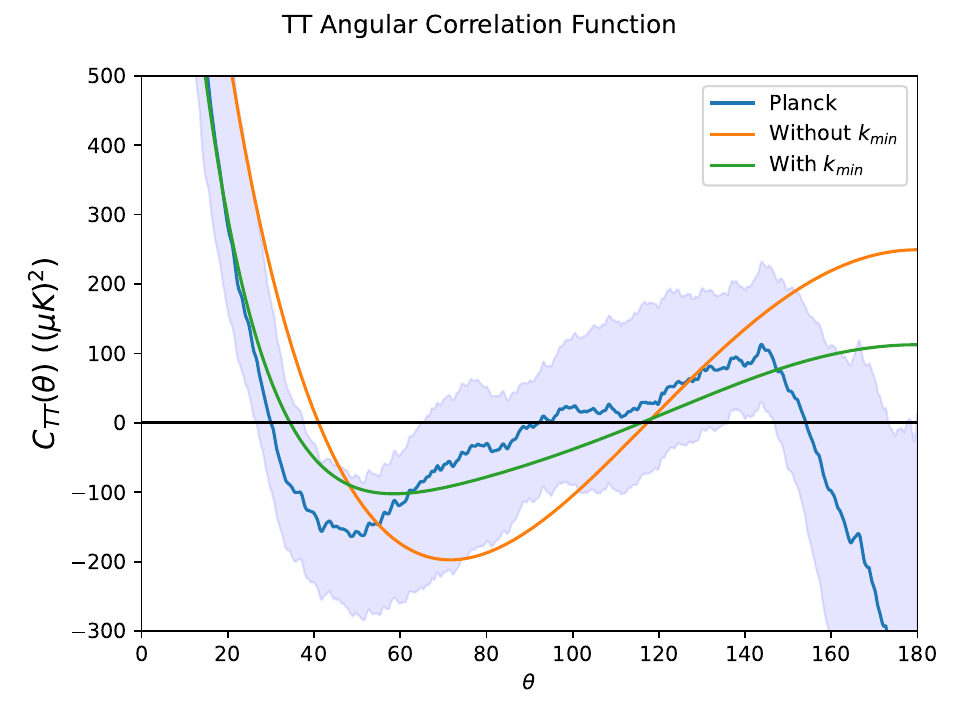}
\caption{TT angular correlation function with (green) and without (red)
$k_{\rm min}$, in comparison with the angular correlation function
calculated from the {\it Planck} data (blue). The shaded region represents
the (non-Gaussian) $1\sigma$ uncertainty obtained from a sample of one 
thousand mock correlation functions generated via Monte Carlo randomization.}
\label{fig:ttctheta}
\end{figure}

The angular power spectrum $C_\ell^{TT}$ is shown in Figure~\ref{fig:ttcl}, and
the corresponding TT angular correlation function is shown in Figure~\ref{fig:ttctheta}. 
To achieve a realistic representation of the {\it Planck} measured angular correlation 
function along with its $1\sigma$ range, we generated a sample of one 
thousand mock realizations of the angular correlation function. More specifically, 
we used the measured $C_\ell^{TT}$ values and their uncertainties, as published in
\cite{PlanckVI:2020}, to generate a thousand sets of mock $C_\ell^{TT}$'s. To address 
the non-Gaussianity of the $C_\ell^{TT}$ errors, we assumed that the upper and 
lower errors in $C_\ell^{TT}$ follow (possibly different) half-Gaussian distributions. 
We then randomized the measured $C_\ell^{TT}$ values within these distributions. For 
each set of mock $C_\ell^{TT}$, we calculated the corresponding angular correlation 
function, thus obtaining a thousand realizations of the mock $C(\theta)$. From this 
sample, we then estimated the non-Gaussian errors using the same approach, i.e.,
allowing the upper and lower error distributions to be half-Gaussians.

\begin{figure}[ht]
\centering
\includegraphics[width=\columnwidth]{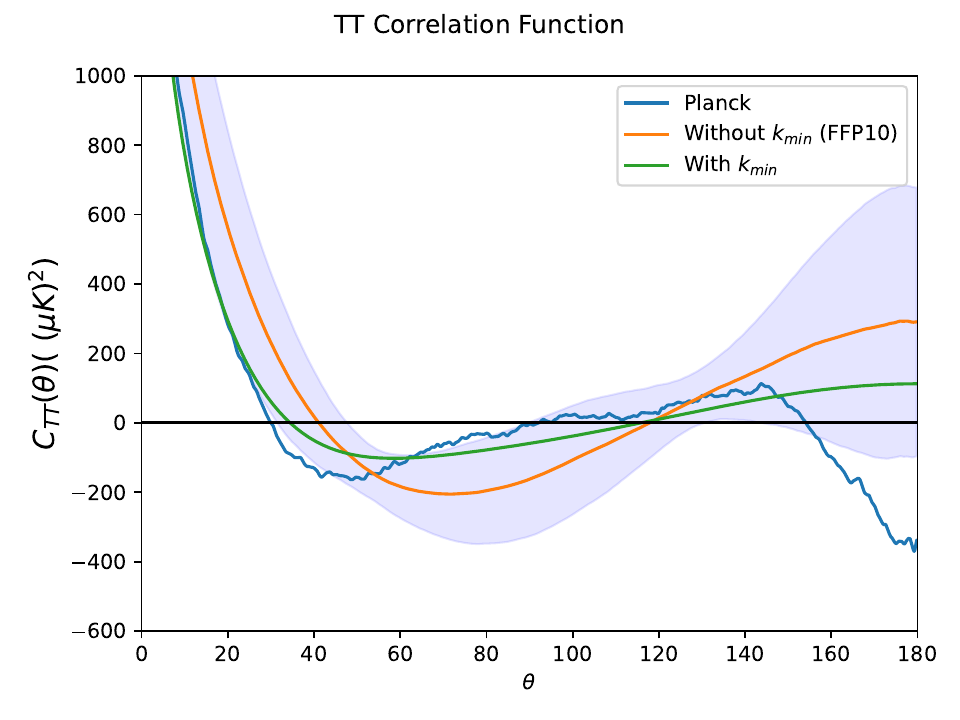}
\caption{Same as Fig.~\ref{fig:ttctheta}, except that here the
shaded region corresponds to the $1\sigma$ cosmic variance, calculated 
utilizing {\it Planck}'s official set of simulations (FFP10) for the
concordance model (red curve).}
\label{fig:ttctheta2}
\end{figure}

The disparity between the predicted and measured angular
correlation functions is sometimes quantified in terms of the cosmic
variance associated with the theoretical calculation. The results in
Figure~\ref{fig:ttctheta} are thus reproduced in Figure~\ref{fig:ttctheta2}, 
where the shaded region now corresponds to this cosmic variance in the
concordance model (red curve), estimated using the {\it Planck} official 
simulations (FFP10 CMB realizations).

These results confirm those obtained earlier for the TT distributions 
\cite{MeliaLopez:2018,Melia:2021b} based on a different methodology 
and the older (i.e., {\it Planck}-2014) data. As demonstrated in 
these previous works, the optimized value of $k_{\rm min}$ 
significantly improves the fits in both Figures~\ref{fig:ttcl} and 
\ref{fig:ttctheta} (and \ref{fig:ttctheta2}) at all angles.

\section{B-mode Polarization}\label{B-mode}
\begin{figure}[ht]
\centering
\includegraphics[width=\columnwidth]{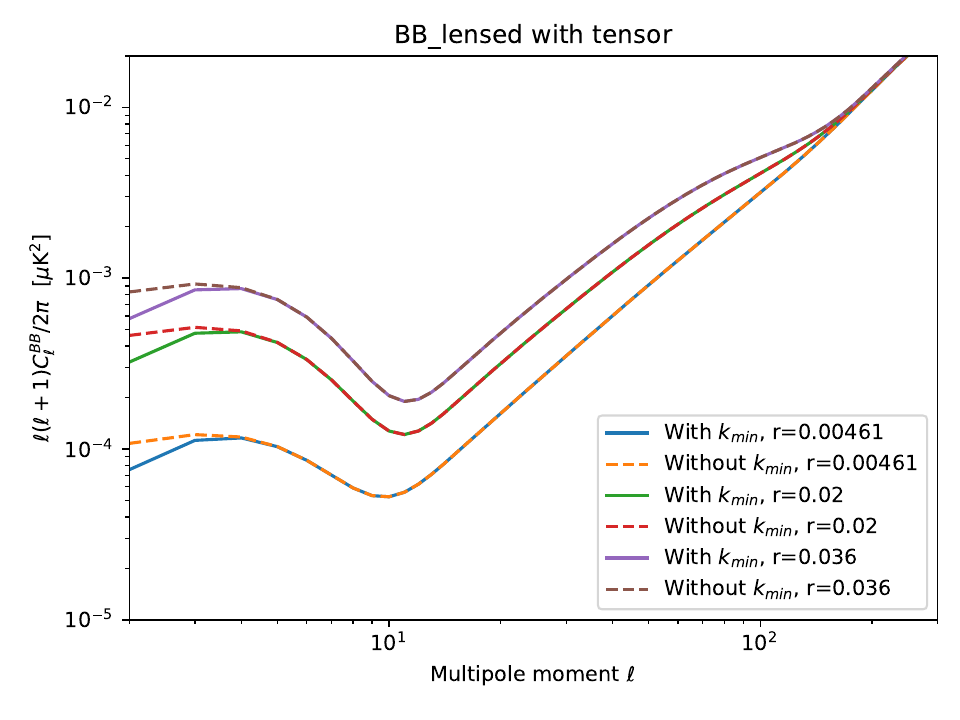}
\caption{BB angular power spectrum for different values of $r$: 
$0.00461$ , $0.02$ and $0.036$ , with (solid) and without (dashed) $k_{\rm min}$.} 
\label{fig:bbclvr}
\end{figure}

The BB angular power spectrum $C_\ell^{BB}$ is calculated assuming a
primordial tensor distribution 

\begin{equation}
P^{(T)}(k) = 
\begin{cases} 
0 & \text{for } k < k_{\rm min}, \\
r \cdot A_{\rm s} \cdot 
\left(\frac{k}{k_{\rm pivot}}\right)^{n_{\rm T}} & \text{for } k \geq k_{\rm min}.
\end{cases}
\label{eq:Pk_t}
\end{equation}
in terms of the tensor-to-scalar ratio $r$ and the spectral index $n_{\rm T}$ of the tensor 
fluctuations, which we assume to be the conventional $n_{\rm T} = -r/8$. The range of values 
discussed here is motivated (i) by the current upper bound $r<0.036$ provided by
{\it Planck} and BICEP2/Keck \cite{Tristram:2021}, and (ii) by the expected 
sensitivity of LiteBIRD, which should allow a measurement to levels less than 
$r=0.004$. It is important to note that B-mode polarization arises, not just 
from these tensor modes in the primordial power spectrum, but also from weak 
lensing effects that convert some of the E-mode polarization into B-mode.
\texttt{CAMB} correctly takes all such effects into consideration
\cite{Hu:2000}. The angular power spectrum $C_\ell^{BB}$ is shown
in Figure \ref{fig:bbclvr}. The BB spectra reveal noticeable 
differences for $\ell\leq 4$, analogously to the situation with TT.
Moreover, the difference increases somewhat with increasing $r$.

\begin{figure*}[ht!]
    \centering
    \begin{subfigure}[b]{0.45\linewidth}
        \includegraphics[width=\linewidth]{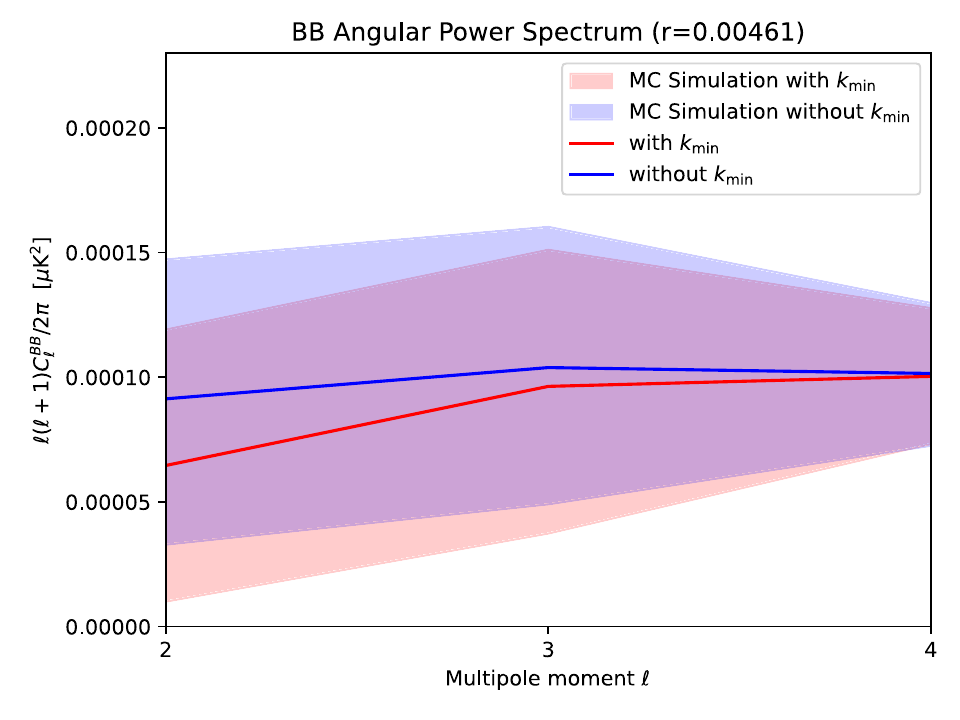}
        \label{fig:sub1}
    \end{subfigure}
    \hfill 
    \begin{subfigure}[b]{0.45\linewidth}
        \includegraphics[width=\linewidth]{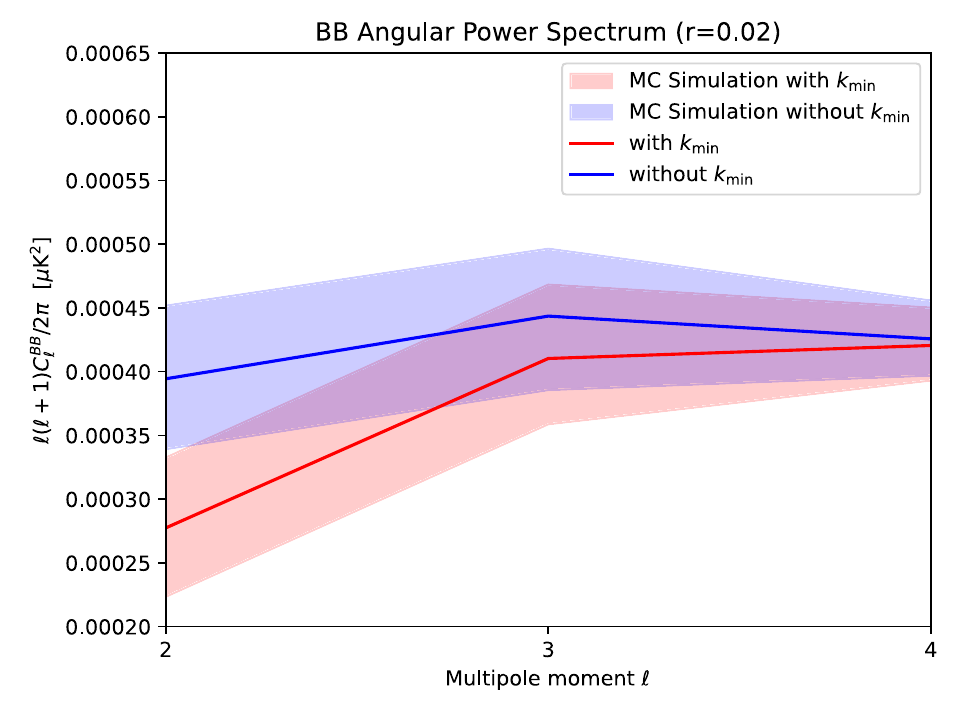}
        \label{fig:sub2}
    \end{subfigure}
    
    \vspace{1em} 
    
    \begin{subfigure}[b]{0.45\linewidth}
        \includegraphics[width=\linewidth]{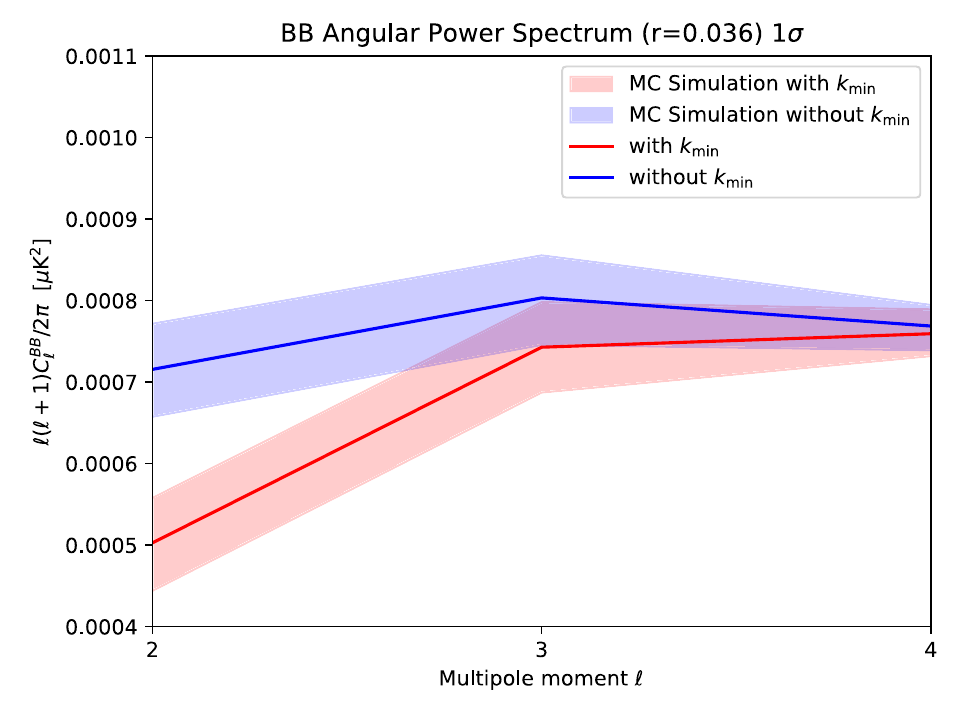}
        \label{fig:sub3}
    \end{subfigure}
    \hfill 
    \begin{subfigure}[b]{0.45\linewidth}
        \includegraphics[width=\linewidth]{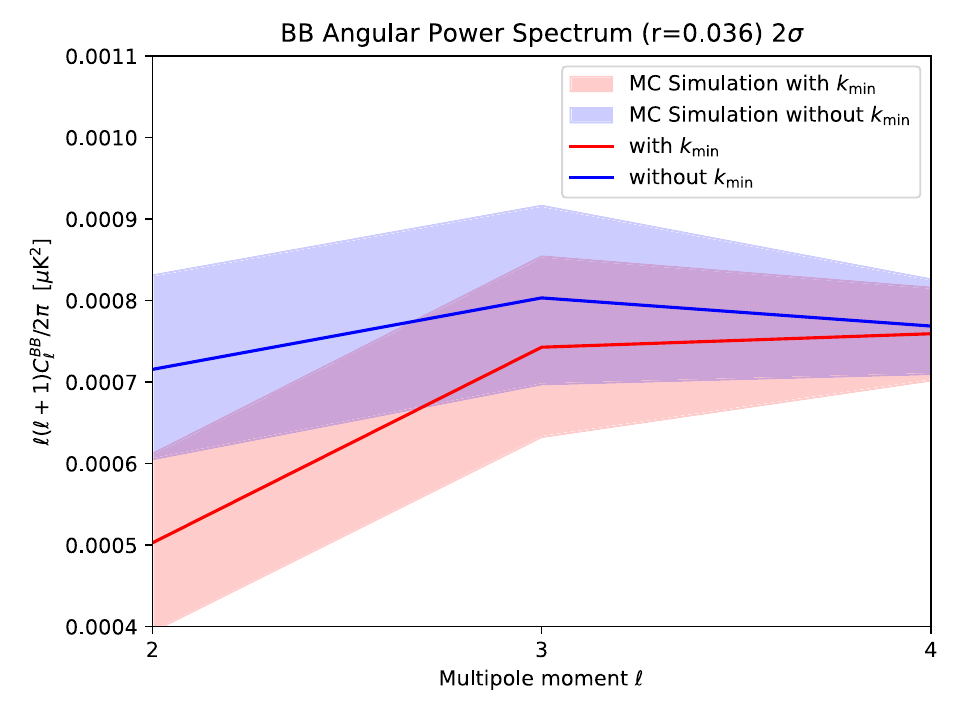}
        \label{fig:sub4}
    \end{subfigure}
    
    \caption{BB angular power spectrum for $\ell\leq 4$. In all four panels, 
the red and blue lines show the results with and without $k_{\rm min}$. The shaded regions 
show the expected total LiteBIRD errors, including foreground residuals. Top-left is for 
$r=0.00461$, with the forecast \(\pm 1 \sigma\) region; Top-right is for $r=0.02$, with 
the expected \(\pm 1 \sigma\) region; Bottom-left is for $r=0.036$, with the forecast 
\(\pm 1 \sigma\) region; Bottom-right is for $r=0.036$, with the expected \(\pm 2 \sigma\)  
region. }
    \label{fig:bbcl}
\end{figure*}

Increasing $r$ mainly affects the low-$\ell$ $C_\ell$'s because of the two 
sources of B-mode polarization in the CMB introduced above: primordial 
gravitational waves (i.e., tensor modes) and weak gravitational lensing.
The influence of these two contributors to $C_\ell^{BB}$ varies in 
significance at different $\ell$'s. Tensor modes mainly affect 
the low-$\ell$ region because they induce B-mode polarization 
primarily on large scales. They generated the B-mode polarization
pattern at photon decoupling ($t\sim 380,000$ years), which 
remained unchanged afterwards. In contrast, weak gravitational 
lensing contributes relatively more to $C_\ell^{BB}$ at higher $\ell$'s. 
The lensing effect warps the null geodesics arriving from the
early Universe, converting some of the E-mode polarization into 
B-mode. This effect is more pronounced on small scales (high 
$\ell$'s), given that lensing depends on the distribution of 
matter on large scales altering the photon paths on smaller 
scales. Thus, when lensing effects are present, tensor modes 
mainly influence the low $\ell$ region of $C_\ell^{BB}$.

The portion of Figure~\ref{fig:bbclvr} most relevant to this analysis
($\ell\leq 4$) is magnified and displayed in the four panels of Figure~\ref{fig:bbcl}.
The difference induced by a nonzero cutoff to $P^{(T)}(k)$ is here compared to 
the expected total uncertainty in the LiteBIRD observations,
believed to be
the most sensitive upcoming B-mode measurements. These simulations begin with
the theoretical curves for $C_\ell^{BB}$, with and without the cutoff. Then,
assuming Gaussian errors for these future measurements (The ``total LiteBIRD 
errors'' we used here were published by the LiteBIRD team \cite{LiteBIRD:2022}. 
For the range of \(\ell\) we are interested in (\(\ell \leq 10\)), the errors 
are presented as Gaussian), we carry out a Monte 
Carlo simulation randomizing the results based on the total (predicted) 
LiteBIRD uncertainty \cite{LiteBIRD:2022} to produce a mock sample of 
possible $C_\ell^{BB}$ distributions as a function of $r$ . We calculate
the $\pm 1\sigma$ and $\pm 2\sigma$ errors assuming a Gaussian distribution
of the $C_\ell^{BB}$ values in the mock sample.

The expected total uncertainty used here is taken directly from the LiteBIRD 
simulations, which includes cosmic variance and the noisy foreground residuals. 
The contributing instrumental systematic uncertainty is considered by the
LiteBIRD team when creating mock input foreground maps for the simulations.

The panels in this figure show the results for three representative values of 
$r$ (the tensor-to-scalar ratio): $0.036$ is the current upper limit of $r$ 
consistent with the latest {\it Planck} data release \cite{PlanckVI:2020}, 
which produces the largest difference between the two curves (with and without 
$k_{\rm min}$); $0.02$ is the smallest $r$ for which the difference between the 
two curves (at $\ell=2$) is still larger than the expected LiteBIRD total 
uncertainty; $0.00461$ coincides with the value utilized by the LiteBIRD team 
when presenting their projected error bars \cite{LiteBIRD:2022}. As noted earlier,
a larger $r$ accentuates the disparity between the cases with and without $k_{\rm min}$.

We can clearly see that the shift introduced by $k_{\rm min}$ is relatively 
insignificant compared to the forecast LiteBIRD total uncertainty when $r$ is 
very small ($r\sim 0.005$). If $r$ is not much smaller than its current upper 
limit, however, the impact of $k_{\rm min}$ should be measurable, especailly at 
$\ell = 2$. The cutoff in $P^{(T)}(k)$ produces an observable signature well 
above cosmic variance and LiteBIRD's expected measurement error. 

\begin{figure}[!htbp]
\centering
\includegraphics[width=\columnwidth]{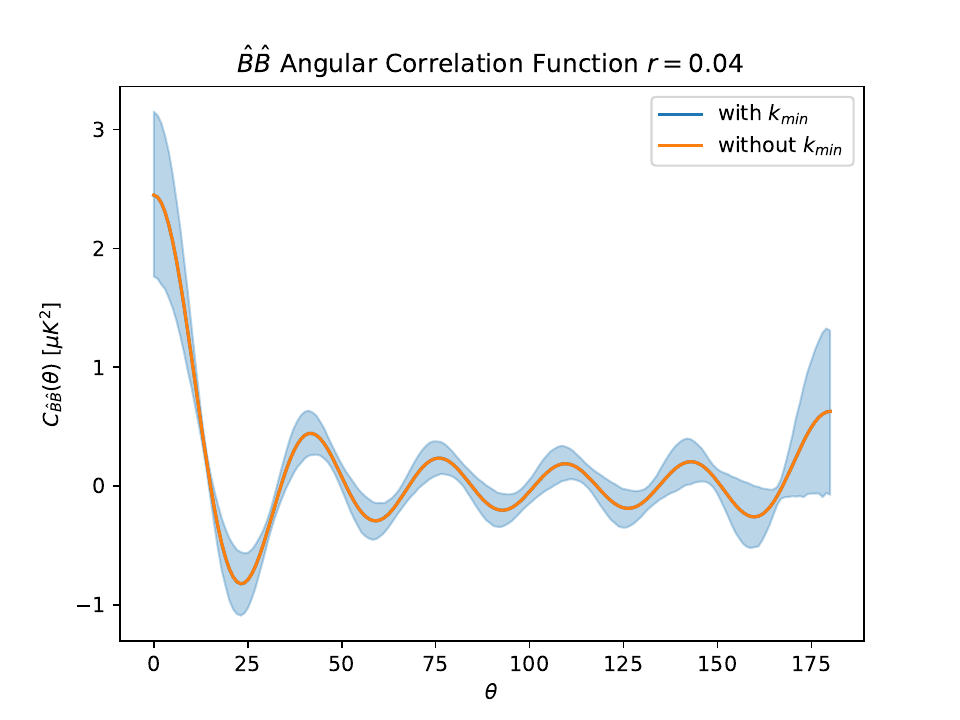}
\caption{$\hat{B}\hat{B}$ angular correlation function for 
$\ell \leq 10$. The plot compares the results with and without 
$k_{\rm min}$. The shaded area represents the $\pm1\sigma$ error
region anticipated by LiteBIRD.}
\label{fig:bbcthetahat2}
\end{figure}

To complete the summary of results from this analysis, we also show in
Figure~\ref{fig:bbcthetahat2} the angular correlation function of the B-mode 
polarization for $r=0.036$. This calculation includes only the range 
$\ell \leq 10$ since only the low-$\ell$ $C_{\ell}$'s may be
differentiated on the basis of their $k_{\rm min}$ value. The $\pm 1\sigma$
uncertainty region corresponding to LiteBIRD's expected measurement
accuracy for $C_{\ell}$ \cite{LiteBIRD:2022} is estimated with a 
ten-thousand step Monte Carlo simulation. The two curves are exceedingly 
close. LiteBIRD may not be able to differentiate between them. As noted
earlier, this outcome is due to the extra factor \(\frac{(\ell+2)!}{(\ell-2)!}\)
appearing in Equation~(\ref{CthetaBB}), which amplifies the dependence of 
the angular correlation function on the higher $\ell$'s. Thus, in spite
of our calculation restricting the range of multipoles to $\ell \leq 10$, 
the impact of $k_{\rm min}$ is largely concealed by the dominance of the
$\ell > 5$ terms.

\section{Conclusion}\label{conclusion}
The hypothesized delayed initiation of inflation would produce a rather
rigid cut-off, $k_{\rm min}$, to both the scalar and tensor primordial
power spectra. In this {\it Letter}, we have sought to find an unambiguous
observational signature of this crucial inflationary feature, which is
manifested in both the TT and BB CMB anisotropies. It is already known
that $k_{\rm min}$ significantly impacts the TT angular correlation
function. But whereas the current {\it Planck} data support the existence
of such a cutoff in the scalar power spectrum, only high-precision 
observations of the B-mode polarization by forthcoming missions can
fully reveal a truncation in both the scalar and tensor modes, thereby 
providing clear, untainted evidence of a delayed de Sitter expansion 
in the early Universe.
 
We have shown that a cutoff $k_{\rm min}$ should be clearly measurable
by LiteBIRD if the tensor-to-scalar ratio, $r$, is not much 
smaller than $\sim 0.02$. At the current upper limit, $r=0.036$, the 
differences in $C_{\ell}^{BB}$ for $\ell < 4$ with and without the 
cutoff easily exceed LiteBIRD's total uncertainty.

The influence of $k_{\rm min}$ on the B-mode angular correlation 
function is less pronounced, due to its greater dependence on the
high-$\ell$ terms in the angular power spectrum. In this context,
the outcomes for B-mode and TT are distincly different. An anomaly
in the $\hat{B}\hat{B}$ angular correlation function is thus
unlikely, again contrasting with the existing situation with TT.

The impact of a measured nonzero cutoff $k_{\rm min}$ in the
B-mode angular power spectrum cannot be overstated. Together with
the existing constraints inferred from TT, this would provide
compelling evidence of a delayed de Sitter expansion in the early
Universe, as required by the new, chaotic inflationary paradigm.



\section*{Acknowledgement} We are very grateful to the anonymous
referees for provinding a detailed, constructive set of 
recommendations that have led to significant improvements in the
presentation of this material.

\bibliographystyle{elsarticle-num-names} 
\bibliography{ms-liu.bib}

\begin{thebibliography}{24}
\expandafter\ifx\csname natexlab\endcsname\relax\def\natexlab#1{#1}\fi
\providecommand{\url}[1]{\texttt{#1}}
\providecommand{\href}[2]{#2}
\providecommand{\path}[1]{#1}
\providecommand{\DOIprefix}{doi:}
\providecommand{\ArXivprefix}{arXiv:}
\providecommand{\URLprefix}{URL: }
\providecommand{\Pubmedprefix}{pmid:}
\providecommand{\doi}[1]{\href{http://dx.doi.org/#1}{\path{#1}}}
\providecommand{\Pubmed}[1]{\href{pmid:#1}{\path{#1}}}
\providecommand{\bibinfo}[2]{#2}
\ifx\xfnm\relax \def\xfnm[#1]{\unskip,\space#1}\fi
\bibitem[{{Starobinski{\v{i}}}(1979)}]{Starobinskii:1979}
\bibinfo{author}{A.~A. {Starobinski{\v{i}}}},
\newblock \bibinfo{title}{{Spectrum of relict gravitational radiation and the
  early state of the universe}},
\newblock \bibinfo{journal}{Soviet Journal of Experimental and Theoretical
  Physics Letters} \bibinfo{volume}{30} (\bibinfo{year}{1979})
  \bibinfo{pages}{682}.
\bibitem[{{Kazanas}(1980)}]{Kazanas:1980}
\bibinfo{author}{D.~{Kazanas}},
\newblock \bibinfo{title}{{Dynamics of the universe and spontaneous symmetry
  breaking}},
\newblock \bibinfo{journal}{\apjl} \bibinfo{volume}{241} (\bibinfo{year}{1980})
  \bibinfo{pages}{L59--L63}. \DOIprefix\doi{10.1086/183361}.
\bibitem[{{Guth}(1981)}]{Guth:1981}
\bibinfo{author}{A.~H. {Guth}},
\newblock \bibinfo{title}{{Inflationary universe: A possible solution to the
  horizon and flatness problems}},
\newblock \bibinfo{journal}{\prd} \bibinfo{volume}{23} (\bibinfo{year}{1981})
  \bibinfo{pages}{347--356}. \DOIprefix\doi{10.1103/PhysRevD.23.347}.
\bibitem[{{Linde}(1982)}]{Linde:1982}
\bibinfo{author}{A.~D. {Linde}},
\newblock \bibinfo{title}{{A new inflationary universe scenario: A possible
  solution of the horizon, flatness, homogeneity, isotropy and primordial
  monopole problems}},
\newblock \bibinfo{journal}{Physics Letters B} \bibinfo{volume}{108}
  (\bibinfo{year}{1982}) \bibinfo{pages}{389--393}.
  \DOIprefix\doi{10.1016/0370-2693(82)91219-9}.
\bibitem[{{Planck Collaboration} et~al.(2014){Planck Collaboration}, {Ade},
  {Aghanim}, {Armitage-Caplan}, {Arnaud}, {Ashdown}, {Atrio-Barandela},
  {Aumont}, {Baccigalupi}, {Banday}, {Barreiro}, and {et al.}}]{Planck:2014}
\bibinfo{author}{{Planck Collaboration}}, \bibinfo{author}{P.~A.~R. {Ade}},
  \bibinfo{author}{N.~{Aghanim}}, \bibinfo{author}{C.~{Armitage-Caplan}},
  \bibinfo{author}{M.~{Arnaud}}, \bibinfo{author}{M.~{Ashdown}},
  \bibinfo{author}{F.~{Atrio-Barandela}}, \bibinfo{author}{J.~{Aumont}},
  \bibinfo{author}{C.~{Baccigalupi}}, \bibinfo{author}{A.~J. {Banday}},
  \bibinfo{author}{R.~B. {Barreiro}}, \bibinfo{author}{{et al.}},
\newblock \bibinfo{title}{{Planck 2013 results. XVI. Cosmological parameters}},
\newblock \bibinfo{journal}{\aap} \bibinfo{volume}{571} (\bibinfo{year}{2014})
  \bibinfo{pages}{A16}. \DOIprefix\doi{10.1051/0004-6361/201321591}.
  \href{http://arxiv.org/abs/1303.5076}{{\tt arXiv:1303.5076}}.
\bibitem[{{Ijjas} et~al.(2013){Ijjas}, {Steinhardt}, and {Loeb}}]{Ijjas:2013}
\bibinfo{author}{A.~{Ijjas}}, \bibinfo{author}{P.~J. {Steinhardt}},
  \bibinfo{author}{A.~{Loeb}},
\newblock \bibinfo{title}{{Inflationary paradigm in trouble after Planck2013}},
\newblock \bibinfo{journal}{Physics Letters B} \bibinfo{volume}{723}
  (\bibinfo{year}{2013}) \bibinfo{pages}{261--266}.
  \DOIprefix\doi{10.1016/j.physletb.2013.05.023}.
  \href{http://arxiv.org/abs/1304.2785}{{\tt arXiv:1304.2785}}.
\bibitem[{{Ijjas} et~al.(2014){Ijjas}, {Steinhardt}, and {Loeb}}]{Ijjas:2014}
\bibinfo{author}{A.~{Ijjas}}, \bibinfo{author}{P.~J. {Steinhardt}},
  \bibinfo{author}{A.~{Loeb}},
\newblock \bibinfo{title}{{Inflationary schism}},
\newblock \bibinfo{journal}{Physics Letters B} \bibinfo{volume}{736}
  (\bibinfo{year}{2014}) \bibinfo{pages}{142--146}.
  \DOIprefix\doi{10.1016/j.physletb.2014.07.012}.
  \href{http://arxiv.org/abs/1402.6980}{{\tt arXiv:1402.6980}}.
\bibitem[{{Hinshaw} et~al.(1996){Hinshaw}, {Branday}, {Bennett}, {Gorski},
  {Kogut}, {Lineweaver}, {Smoot}, and {Wright}}]{Hinshaw:1996}
\bibinfo{author}{G.~{Hinshaw}}, \bibinfo{author}{A.~J. {Branday}},
  \bibinfo{author}{C.~L. {Bennett}}, \bibinfo{author}{K.~M. {Gorski}},
  \bibinfo{author}{A.~{Kogut}}, \bibinfo{author}{C.~H. {Lineweaver}},
  \bibinfo{author}{G.~F. {Smoot}}, \bibinfo{author}{E.~L. {Wright}},
\newblock \bibinfo{title}{{Two-Point Correlations in the COBE DMR Four-Year
  Anisotropy Maps}},
\newblock \bibinfo{journal}{\apjl} \bibinfo{volume}{464} (\bibinfo{year}{1996})
  \bibinfo{pages}{L25}. \DOIprefix\doi{10.1086/310076}.
  \href{http://arxiv.org/abs/astro-ph/9601061}{{\tt arXiv:astro-ph/9601061}}.
\bibitem[{{Bennett} et~al.(2003){Bennett}, {Hill}, {Hinshaw}, {Nolta},
  {Odegard}, {Page}, {Spergel}, {Weiland}, {Wright}, {Halpern}, {Jarosik},
  {Kogut}, {Limon}, {Meyer}, {Tucker}, and {Wollack}}]{Bennett:2003}
\bibinfo{author}{C.~L. {Bennett}}, \bibinfo{author}{R.~S. {Hill}},
  \bibinfo{author}{G.~{Hinshaw}}, \bibinfo{author}{M.~R. {Nolta}},
  \bibinfo{author}{N.~{Odegard}}, \bibinfo{author}{L.~{Page}},
  \bibinfo{author}{D.~N. {Spergel}}, \bibinfo{author}{J.~L. {Weiland}},
  \bibinfo{author}{E.~L. {Wright}}, \bibinfo{author}{M.~{Halpern}},
  \bibinfo{author}{N.~{Jarosik}}, \bibinfo{author}{A.~{Kogut}},
  \bibinfo{author}{M.~{Limon}}, \bibinfo{author}{S.~S. {Meyer}},
  \bibinfo{author}{G.~S. {Tucker}}, \bibinfo{author}{E.~{Wollack}},
\newblock \bibinfo{title}{{First-Year Wilkinson Microwave Anisotropy Probe
  (WMAP) Observations: Foreground Emission}},
\newblock \bibinfo{journal}{\apjs} \bibinfo{volume}{148} (\bibinfo{year}{2003})
  \bibinfo{pages}{97--117}. \DOIprefix\doi{10.1086/377252}.
  \href{http://arxiv.org/abs/astro-ph/0302208}{{\tt arXiv:astro-ph/0302208}}.
\bibitem[{{Planck Collaboration} et~al.(2020){Planck Collaboration}, {Aghanim},
  {Akrami}, {Ashdown}, {Aumont}, {Baccigalupi}, {Ballardini}, {Banday},
  {Barreiro}, {Bartolo}, {Basak}, and {et al.}}]{PlanckVI:2020}
\bibinfo{author}{{Planck Collaboration}}, \bibinfo{author}{N.~{Aghanim}},
  \bibinfo{author}{Y.~{Akrami}}, \bibinfo{author}{p.~M. {Ashdown}},
  \bibinfo{author}{J.~{Aumont}}, \bibinfo{author}{C.~{Baccigalupi}},
  \bibinfo{author}{M.~{Ballardini}}, \bibinfo{author}{A.~J. {Banday}},
  \bibinfo{author}{R.~B. {Barreiro}}, \bibinfo{author}{N.~{Bartolo}},
  \bibinfo{author}{S.~{Basak}}, \bibinfo{author}{{et al.}},
\newblock \bibinfo{title}{{Planck 2018 results. VI. Cosmological parameters}},
\newblock \bibinfo{journal}{\aap} \bibinfo{volume}{641} (\bibinfo{year}{2020})
  \bibinfo{pages}{A6}. \DOIprefix\doi{10.1051/0004-6361/201833910}.
  \href{http://arxiv.org/abs/1807.06209}{{\tt arXiv:1807.06209}}.
\bibitem[{{Copi} et~al.(2009){Copi}, {Huterer}, {Schwarz}, and
  {Starkman}}]{Copi:2009}
\bibinfo{author}{C.~J. {Copi}}, \bibinfo{author}{D.~{Huterer}},
  \bibinfo{author}{D.~J. {Schwarz}}, \bibinfo{author}{G.~D. {Starkman}},
\newblock \bibinfo{title}{{No large-angle correlations on the non-Galactic
  microwave sky}},
\newblock \bibinfo{journal}{\mnras} \bibinfo{volume}{399}
  (\bibinfo{year}{2009}) \bibinfo{pages}{295--303}.
  \DOIprefix\doi{10.1111/j.1365-2966.2009.15270.x}.
  \href{http://arxiv.org/abs/0808.3767}{{\tt arXiv:0808.3767}}.
\bibitem[{{Melia} and {L{\'o}pez-Corredoira}(2018)}]{MeliaLopez:2018}
\bibinfo{author}{F.~{Melia}}, \bibinfo{author}{M.~{L{\'o}pez-Corredoira}},
\newblock \bibinfo{title}{{Evidence of a truncated spectrum in the angular
  correlation function of the cosmic microwave background}},
\newblock \bibinfo{journal}{\aap} \bibinfo{volume}{610} (\bibinfo{year}{2018})
  \bibinfo{pages}{A87}. \DOIprefix\doi{10.1051/0004-6361/201732181}.
  \href{http://arxiv.org/abs/1712.07847}{{\tt arXiv:1712.07847}}.
\bibitem[{{Liu} and {Melia}(2020)}]{LiuMelia:2020}
\bibinfo{author}{J.~{Liu}}, \bibinfo{author}{F.~{Melia}},
\newblock \bibinfo{title}{{Viability of slow-roll inflation in light of the
  non-zero k$_{min}$ measured in the cosmic microwave background power
  spectrum}},
\newblock \bibinfo{journal}{Proceedings of the Royal Society of London Series
  A} \bibinfo{volume}{476} (\bibinfo{year}{2020}) \bibinfo{pages}{20200364}.
  \DOIprefix\doi{10.1098/rspa.2020.0364}.
  \href{http://arxiv.org/abs/2006.02510}{{\tt arXiv:2006.02510}}.
\bibitem[{{Lewis} et~al.(2000){Lewis}, {Challinor}, and {Lasenby}}]{Lewis:2000}
\bibinfo{author}{A.~{Lewis}}, \bibinfo{author}{A.~{Challinor}},
  \bibinfo{author}{A.~{Lasenby}},
\newblock \bibinfo{title}{{Efficient Computation of Cosmic Microwave Background
  Anisotropies in Closed Friedmann-Robertson-Walker Models}},
\newblock \bibinfo{journal}{\apj} \bibinfo{volume}{538} (\bibinfo{year}{2000})
  \bibinfo{pages}{473--476}. \DOIprefix\doi{10.1086/309179}.
  \href{http://arxiv.org/abs/astro-ph/9911177}{{\tt arXiv:astro-ph/9911177}}.
\bibitem[{{LiteBIRD Collaboration} et~al.(2023){LiteBIRD Collaboration},
  {Allys}, {Arnold}, {Aumont}, {Aurlien}, {Azzoni}, {Baccigalupi}, {Banday},
  {Banerji}, {Barreiro}, {Bartolo}, {Bautista}, and {et al.}}]{LiteBIRD:2022}
\bibinfo{author}{{LiteBIRD Collaboration}}, \bibinfo{author}{E.~{Allys}},
  \bibinfo{author}{K.~{Arnold}}, \bibinfo{author}{J.~{Aumont}},
  \bibinfo{author}{R.~{Aurlien}}, \bibinfo{author}{S.~{Azzoni}},
  \bibinfo{author}{C.~{Baccigalupi}}, \bibinfo{author}{A.~J. {Banday}},
  \bibinfo{author}{R.~{Banerji}}, \bibinfo{author}{R.~B. {Barreiro}},
  \bibinfo{author}{N.~{Bartolo}}, \bibinfo{author}{L.~{Bautista}},
  \bibinfo{author}{{et al.}},
\newblock \bibinfo{title}{{Probing cosmic inflation with the LiteBIRD cosmic
  microwave background polarization survey}},
\newblock \bibinfo{journal}{Progress of Theoretical and Experimental Physics}
  \bibinfo{volume}{2023} (\bibinfo{year}{2023}) \bibinfo{pages}{042F01}.
  \DOIprefix\doi{10.1093/ptep/ptac150}.
  \href{http://arxiv.org/abs/2202.02773}{{\tt arXiv:2202.02773}}.
\bibitem[{{Bardeen} et~al.(1983){Bardeen}, {Steinhardt}, and
  {Turner}}]{Bardeen:1983}
\bibinfo{author}{J.~M. {Bardeen}}, \bibinfo{author}{P.~J. {Steinhardt}},
  \bibinfo{author}{M.~S. {Turner}},
\newblock \bibinfo{title}{{Spontaneous creation of almost scale-free density
  perturbations in an inflationary universe}},
\newblock \bibinfo{journal}{\prd} \bibinfo{volume}{28} (\bibinfo{year}{1983})
  \bibinfo{pages}{679--693}. \DOIprefix\doi{10.1103/PhysRevD.28.679}.
\bibitem[{{Zaldarriaga} and {Seljak}(1997)}]{Zaldarriaga:1997}
\bibinfo{author}{M.~{Zaldarriaga}}, \bibinfo{author}{U.~{Seljak}},
\newblock \bibinfo{title}{{All-sky analysis of polarization in the microwave
  background}},
\newblock \bibinfo{journal}{\prd} \bibinfo{volume}{55} (\bibinfo{year}{1997})
  \bibinfo{pages}{1830--1840}. \DOIprefix\doi{10.1103/PhysRevD.55.1830}.
  \href{http://arxiv.org/abs/astro-ph/9609170}{{\tt arXiv:astro-ph/9609170}}.
\bibitem[{{Kosowsky}(1996)}]{Kosowsky:1996}
\bibinfo{author}{A.~{Kosowsky}},
\newblock \bibinfo{title}{{Cosmic microwave background polarization.}},
\newblock \bibinfo{journal}{Annals of Physics} \bibinfo{volume}{246}
  (\bibinfo{year}{1996}) \bibinfo{pages}{49--85}.
  \DOIprefix\doi{10.1006/aphy.1996.0020}.
  \href{http://arxiv.org/abs/astro-ph/9501045}{{\tt arXiv:astro-ph/9501045}}.
\bibitem[{{Polnarev}(1985)}]{Polnarev:1985}
\bibinfo{author}{A.~G. {Polnarev}},
\newblock \bibinfo{title}{{Polarization and Anisotropy Induced in the Microwave
  Background by Cosmological Gravitational Waves}},
\newblock \bibinfo{journal}{\sovast} \bibinfo{volume}{29}
  (\bibinfo{year}{1985}) \bibinfo{pages}{607--613}.
\bibitem[{{Yoho} et~al.(2015){Yoho}, {Aiola}, {Copi}, {Kosowsky}, and
  {Starkman}}]{Yoho:2015}
\bibinfo{author}{A.~{Yoho}}, \bibinfo{author}{S.~{Aiola}},
  \bibinfo{author}{C.~J. {Copi}}, \bibinfo{author}{A.~{Kosowsky}},
  \bibinfo{author}{G.~D. {Starkman}},
\newblock \bibinfo{title}{{Microwave background polarization as a probe of
  large-angle correlations}},
\newblock \bibinfo{journal}{\prd} \bibinfo{volume}{91} (\bibinfo{year}{2015})
  \bibinfo{pages}{123504}. \DOIprefix\doi{10.1103/PhysRevD.91.123504}.
  \href{http://arxiv.org/abs/1503.05928}{{\tt arXiv:1503.05928}}.
\bibitem[{{Melia} et~al.(2021){Melia}, {Ma}, {Wei}, and {Yu}}]{Melia:2021b}
\bibinfo{author}{F.~{Melia}}, \bibinfo{author}{Q.~{Ma}}, \bibinfo{author}{J.-J.
  {Wei}}, \bibinfo{author}{B.~{Yu}},
\newblock \bibinfo{title}{{Hint of a truncated primordial spectrum from the CMB
  large-scale anomalies}},
\newblock \bibinfo{journal}{\aap} \bibinfo{volume}{655} (\bibinfo{year}{2021})
  \bibinfo{pages}{A70}. \DOIprefix\doi{10.1051/0004-6361/202141251}.
  \href{http://arxiv.org/abs/2109.05480}{{\tt arXiv:2109.05480}}.
\bibitem[{{Liu} and {Melia}(2024)}]{LiuMelia:2024}
\bibinfo{author}{J.~{Liu}}, \bibinfo{author}{F.~{Melia}},
\newblock \bibinfo{title}{{Challenges to Inflation in the post-Planck Era}},
\newblock \bibinfo{journal}{\apj} \bibinfo{volume}{In press}
  (\bibinfo{year}{2024}).
\bibitem[{{Tristram} et~al.(2021){Tristram}, {Banday}, {G{\'o}rski},
  {Keskitalo}, {Lawrence}, {Andersen}, {Barreiro}, {Borrill}, {Eriksen},
  {Fernandez-Cobos}, {Kisner}, {Mart{\'\i}nez-Gonz{\'a}lez}, {Partridge},
  {Scott}, {Svalheim}, {Thommesen}, and {Wehus}}]{Tristram:2021}
\bibinfo{author}{M.~{Tristram}}, \bibinfo{author}{A.~J. {Banday}},
  \bibinfo{author}{K.~M. {G{\'o}rski}}, \bibinfo{author}{R.~{Keskitalo}},
  \bibinfo{author}{C.~R. {Lawrence}}, \bibinfo{author}{K.~J. {Andersen}},
  \bibinfo{author}{R.~B. {Barreiro}}, \bibinfo{author}{J.~{Borrill}},
  \bibinfo{author}{H.~K. {Eriksen}}, \bibinfo{author}{R.~{Fernandez-Cobos}},
  \bibinfo{author}{T.~S. {Kisner}},
  \bibinfo{author}{E.~{Mart{\'\i}nez-Gonz{\'a}lez}},
  \bibinfo{author}{B.~{Partridge}}, \bibinfo{author}{D.~{Scott}},
  \bibinfo{author}{T.~L. {Svalheim}}, \bibinfo{author}{H.~{Thommesen}},
  \bibinfo{author}{I.~K. {Wehus}},
\newblock \bibinfo{title}{{Planck constraints on the tensor-to-scalar ratio}},
\newblock \bibinfo{journal}{\aap} \bibinfo{volume}{647} (\bibinfo{year}{2021})
  \bibinfo{pages}{A128}. \DOIprefix\doi{10.1051/0004-6361/202039585}.
  \href{http://arxiv.org/abs/2010.01139}{{\tt arXiv:2010.01139}}.
\bibitem[{{Hu}(2000)}]{Hu:2000}
\bibinfo{author}{W.~{Hu}},
\newblock \bibinfo{title}{{Weak lensing of the CMB: A harmonic approach}},
\newblock \bibinfo{journal}{\prd} \bibinfo{volume}{62} (\bibinfo{year}{2000})
  \bibinfo{pages}{043007}. \DOIprefix\doi{10.1103/PhysRevD.62.043007}.
  \href{http://arxiv.org/abs/astro-ph/0001303}{{\tt arXiv:astro-ph/0001303}}.

\end{thebibliography}





\end{document}